\documentclass[prd,singlecolumn]{revtex4}
\pdfoutput=1
\usepackage{amssymb,latexsym}
\usepackage{amsmath,amsbsy,bbm}
\usepackage{ifpdf}
\usepackage{epsfig,bm}
\usepackage{graphicx,comment}
\usepackage{color}
\usepackage{soul}
\usepackage{mathtools}
\usepackage{comment}
\usepackage[normalem]{ulem}
\begin{document}

\title{A Machine Learning study of the two-dimensional antiferromagnetic $Q$-state Potts model on the square lattice }
\author{Shang-Wei Li}
\affiliation{Department of Physics, National Taiwan Normal University,
  88, Sec.4, Ting-Chou Rd., Taipei 116, Taiwan}
\author{Kai-Wei Huang}
\affiliation{Department of Physics, National Taiwan Normal University,
	88, Sec.4, Ting-Chou Rd., Taipei 116, Taiwan}
\author{Chien-Ting Chen}
\affiliation{Department of Physics, National Taiwan Normal University,
	88, Sec.4, Ting-Chou Rd., Taipei 116, Taiwan}	
\author{Fu-Jiun Jiang}
\email[]{fjjiang@ntnu.edu.tw}
\affiliation{Department of Physics, National Taiwan Normal University,
88, Sec.4, Ting-Chou Rd., Taipei 116, Taiwan}


\begin{abstract}

The critical phenomena of 2D antiferromagnetic $Q$-state Potts model on the square lattice with $Q=2,3,4,5$ and 6 are investigated using supervised neural networks (NN). Conventional
	supervised and unsupervised NNs either fail or have difficulty in identifying
	the correct critical behaviors of these models. Remarkably, despite the fact that the NNs employed here are trained with a particular fixed type training set which contains no information about the considered models (except for $Q=2$), 
	they correctly compute the critical temperatures for the studied physical systems.
	Particularly, the NN outcomes suggest convincingly that the $Q=3$ model is critical only at zero temperature and $Q=4,5,6$ models remain
	disordered at all temperatures. Apart from the studied models, our NNs trained with exactly the same training set have successfully determined the critical temperature(s) and the nature of the associated phase transition(s) of another physical system differing dramatically from those considered here. Therefore, our NN has the potential to be a universally applicable NN model in studying phase transitions.
\end{abstract}

\maketitle

\section{Introduction}

Recently, the ideas of Machine Learning techniques have broadened their applications impressively. The usage of these advanced methods is found those days in basically every fields of science, from fundamental research to industry applications. One such scientific field is the physics. In particular, neural networks (NN) are employed to study critical phenomena of many physical models \cite{Tan16,Car16,Nie16,Kim18,Li18,Chn18,Car19,Meh19,Don19,Fuk21,Tir22,Che23,Suk24,Che25}.

Indeed, to study the critical behavior (phase transition) of a physical system with the conventional order-parameter approach, the information of the associated Hamiltonian and how the ground state of the considered system is ordered should be available in advance before one can perform the investigation. Assuming a black box spin configurations is given from an experiment of a unknown
system, then it is a challenge to study the associated critical phenomenon using the order-parameter approach. Under such situation, the NN method can be employed to uncover the related phase transition since NN can detect the change of the spin patterns before and after the transition, hence locates the critical point without any microscopic details of the unknown system. It should also be pointed out that rigor of the conventional order-parameter approach has been well-established, therefore, the NN and order-parameter methods are complementary to each other.   

Potts models are extensions of the Ising model \cite{Pot52,Dom77,Ban80,Gre81,Sai82,Wu82,Ono86,Wan89,Wan90}. Instead of taking only two possible values 1 and -1 as that in Ising model, the Potts variables can be $Q$ discrete consecutive positive integers, namely, 1, 2, 3,...$q-2$, $q-1$, and $Q$. Due to this, the Potts models have very rich critical phenomena in various geometries and dimensions, see Ref.~\cite{Wu82} for the details.

Depending on the sign of the interactions between any pairs of nearest Potts spins, the $Q$-state Potts model can either be a ferromagnetic or an antiferromagnetic system. The antiferromagnetic Potts models are more challenge to study than their ferromagnetic counterparts due to the fact that the antiferromagnetic systems have extremely highly degenerated ground states \cite{Wu82}. Consequently, investigating antiferromagnetic Potts model using NN methods requires huge amount of computing effort. Apart from the computational complexity, the critical phenomena of ferromagnetic and antiferromagnetic Potts models are quite different as well. On two-dimensional (2D) square lattice and $q \ge 2$, while the critical temperatures $T_c$ for $Q$-state ferromagnetic Potts models exist and fulfill $T_c > 0$, only $Q=2$ ($T_c = 1.1346$) and $Q=3$ ($T_c = 0$) antiferromagnetic Potts models can be ordered \cite{Wu82,Sal98}. Because of these untypical features, conventional
supervised and unsupervised NNs, which consider real physical quantities as the training sets, either fail or have difficulty in identifying the correct critical behaviors of the 2D antiferromagnetic $Q$-Potts model for $q \ge 3$ \cite{Tse25}.

While compared to the traditional order-parameter approach, the use of the NN techniques have the advantage of requiring no information of how the system is ordered, to train a NN with real physical configurations,
like those typically done in the literature \cite{Car16},
take a lot of time. In particular, when different models or linear system sizes $L$ are considered, one needs to train a new NN that is suitable to study the targeted new model with the given $L$. Moreover, performing the NN predictions needs quite amount of storage space as well.
Due to these disadvantages mentioned above, to promote the widely adoption of NN approach in reality, a much more efficient NN procedure is highly desirable.

Due to the unusual critical characteristics of the 2D antiferromagnetic Potts model, to our knowledge, no NN investigation for these models are available except 
the one given in Ref.~\cite{Tse25}. The NN computations performed in Ref.~\cite{Tse25} 
employ two ferromagnetic types artificially made configurations as the training set.
In this study, instead of using real spin configurations to train a NN,
we follow a similar idea of Ref.~\cite{Tan20.1} to conduct the training procedure. Specifically, on a square area, two artificially made configurations for which the associated spins have their values in 1 and -1 alternatively in both the rows and columns are used as the building unit of training set. This method of constructing the training set is used in Ref.~\cite{Li25} to study the critical behaviors of
2D antiferromagnetic Ising model with multi-interactions on the triangular lattice.  

Remarkably, by elegantly converting the Potts spins into Ising spins, the NN constructed here, namely an extremely simple multilayer perceptron (MLP), which is trained with two artificially made stagger-type configurations is capable of computing the critical behaviors correctly for the antiferromagnetic Potts models considered in this study. In particular, the MLP outcomes clearly demonstrate that a certain order is established for $Q=3$ model close to zero temperature. Moreover, evidence showing that $Q=4,5$ and 6 models are always disordered at any temperatures including the zero temperature is revealed from the MLP results as well. 

In Ref.~\cite{Li25}, the MLP used in this study has also been considered to uncover the critical phenomena of 2D antiferromagnetic Ising model with nearest and next-nearest interactions successfully. Hence, this MLP may be applicable to other complicated systems than the models studied here and in Ref.~\cite{Li25}. In other words, it has the potential to overcome the
disadvantages of NN methods, namely, tedious retraining when various models are studied and the requirement of huge storage capacity, as discussed above. In particular, even when only two MLPs are used, each of them identifies the correct critical behaviors of
$Q=3,4,5,6$ antiferromagnetic Potts models with various linear system sizes. The
critical temperatures and the nature of the phase transitions for the 4- and 8-state ferromagnetic Potts models are determined
accurately and correctly with one of the two aforementioned MLPs as well.

It should be pointed out that the ground-state configurations of the 2D antiferromagnetic Ising model with nearest and next-nearest interactions, the 2D antiferromagnetic Potts model, and
the 2D ferromagnetic Potts model are distinct from each other dramatically. Yet, a simple MLP trained with the introduced artificially made training set can calculate the corresponding critical temperatures precisely and can determine the nature of the associated phase transitions without ambiguity. To summarize, in this study, the use of the same MLP as the one considered in Ref.~\cite{Li25} is to examine the universal applicability of the MLP, not just a repetition of the established NN architecture.

The same MLP but trained with ferromagnet-like configurations are shown to be universal, i.e. it can compute the critical points and even the critical exponents of many models differing significantly among themselves \cite{Tse22,Pen23,Tse23,Tse241,Tse242,Jia24,Tse25}. It will be intriguing to examine whether this is the case for current MLP.

The rest of the paper is organized as follows. After the introduction, the studied model, the employed NN, the associated
training and testing sets, and the considered quantities associated with NN are described in detail in Sect. II and III. Then the NN outcomes are 
presented in Sect. IV. In particular, the evidence demonstrating
the fact that the used NN can identify the critical phenomena of the studied models correctly is shown. Finally, we conclude this investigation in Sect. V.

\section{The considered model}

The Hamiltonian $H$ of the 2D antiferromagnetic $Q$-state Potts model on the square lattice considered here can be expressed as \cite{Wu82,Wan89,Wan90}
\begin{equation}
	\beta H = \beta \sum_{\left< ij\right>} \delta_{\sigma_i,\sigma_j},
	\label{eqn}
\end{equation}
where $\beta$ is the inverse temperature ($1/T$), $\delta$ stands for the Kronecker function ($\delta_{ab} = 1$ for $a = b$, and $\delta_{cd} = 0$ if
$c \neq d$), and the Potts variable
$\sigma_i$ at each site $i$ takes an integer value from $\{1,2,...,q-1,q\}$.

\section{The employed NN}

The NN used in this investigation is a multilayer perceptron (ML) and has been described in detail in Ref.~\cite{Li25}. For completeness, in this study we introduce this MLP for the benefit of the readers. As a result,
the NN-related sections demonstrated here are (largely) overlapping with the related materials of Ref.~\cite{Li25} since in
principle the same MLP is used in both here and in Ref.~\cite{Li25}.

\begin{figure}
	\includegraphics[width=0.7\columnwidth]{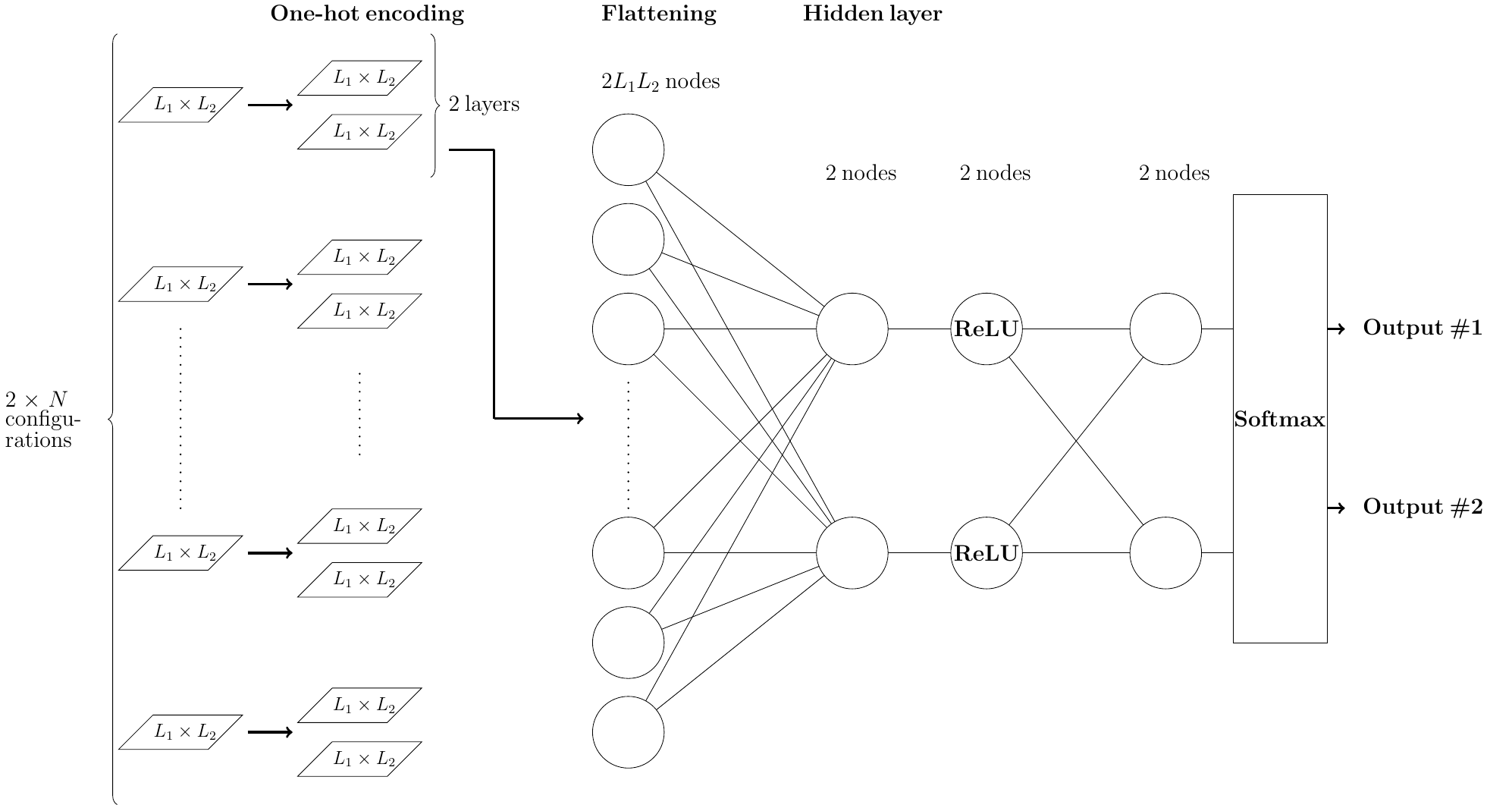}
	\caption{A graphical representation for one of the MLP considered in this study. Here, the value of $N$, i.e. the number of each type of configurations included in the whole training set, is either 200 or 400. The figure is taken from Ref.~\cite{Li25}.}
	\label{nn}
\end{figure}

The NN considered here (and in Ref.~\cite{Li25}) is a simple MLP which consists of one input layer, one hidden layer of two neurons or any other number of neurons, and one output layer. Fig.~\ref{nn} is the pictorial
representation of one of the employed NNs (The figure is taken from Ref.~\cite{Li25}).
Similar to the standard training strategy, the activations (nonlinear) functions used
for the hidden layer and the output layer are
ReLU and Softmax, respectively. The procedure of one-hot encoding is considered
between the input and hidden layers. $L_2$ regularization with the input parameter being 1 is considered as well to avoid overfitting. The algorithm and optimizer employed are minibatch and adam, respectively. Finally,
the MLP outputs are two-component vectors.

\subsection{Training procedure}

\begin{figure}
	\includegraphics[width=0.5\columnwidth]{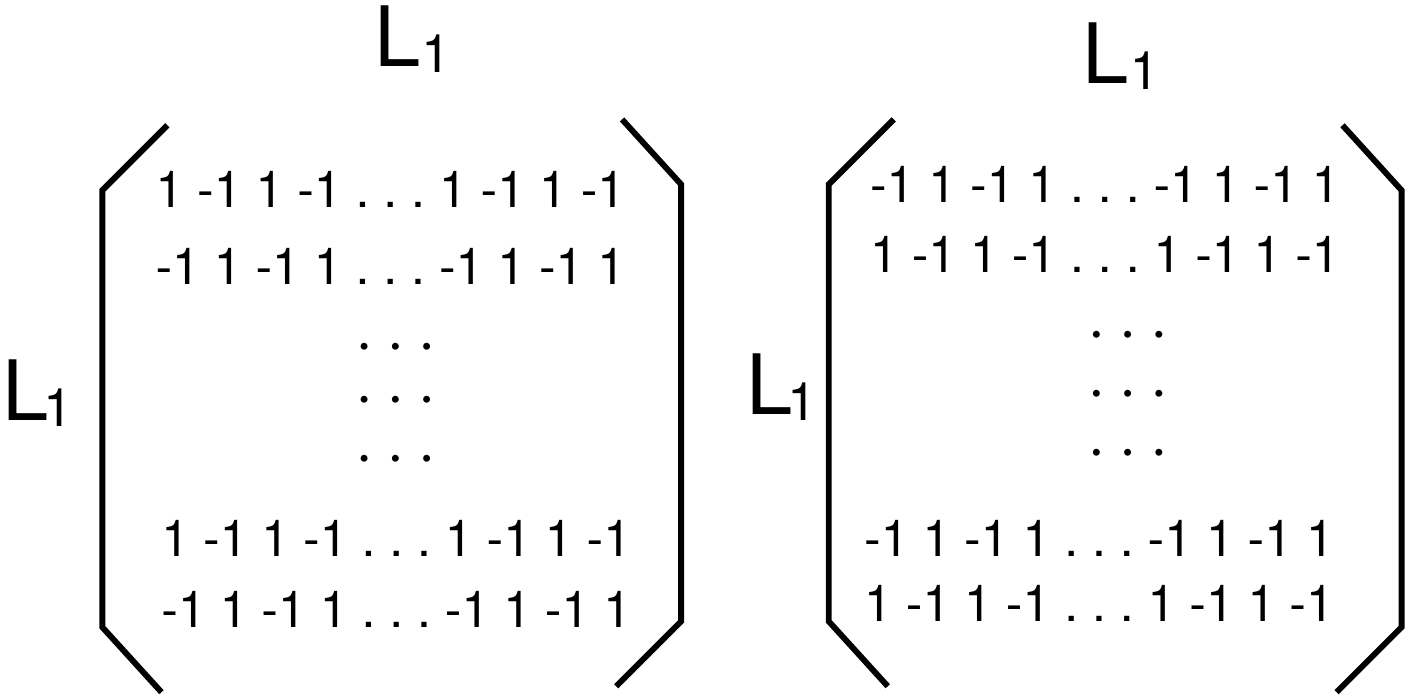}
	\caption{The building units of the training set employed in this study. The value of $L_1$ can be varied. The figure is taken from Ref.~\cite{Li25}. }
	\label{train}
\end{figure}

Instead of using real spin configurations obtained from the Monte Carlo simulations, two kinds of artificially made (stagger-types) configurations are considered as the training set. Fig.~\ref{train} shows the two building units of the training set employed in this study (The figure is from Ref.~\cite{Li25}). Specifically,
the training set consists of $N$ ($N=200$ or 400 in this study) copies of both the left and the right panels of fig.~\ref{train}. Here each building unit is made up of $L_1$ by $L_1$ sites and the value of $L_1$ can be varied. Moreover, the MLP output labels for the training configurations resulting from the left and the right panels of fig.~\ref{train} are $(1,0)$ and $(0,1)$, respectively. Finally, we use 800 to 1600 epochs for the trainings. 

\subsection{The construction of testing set}

\begin{figure}
	\includegraphics[width=0.4\columnwidth]{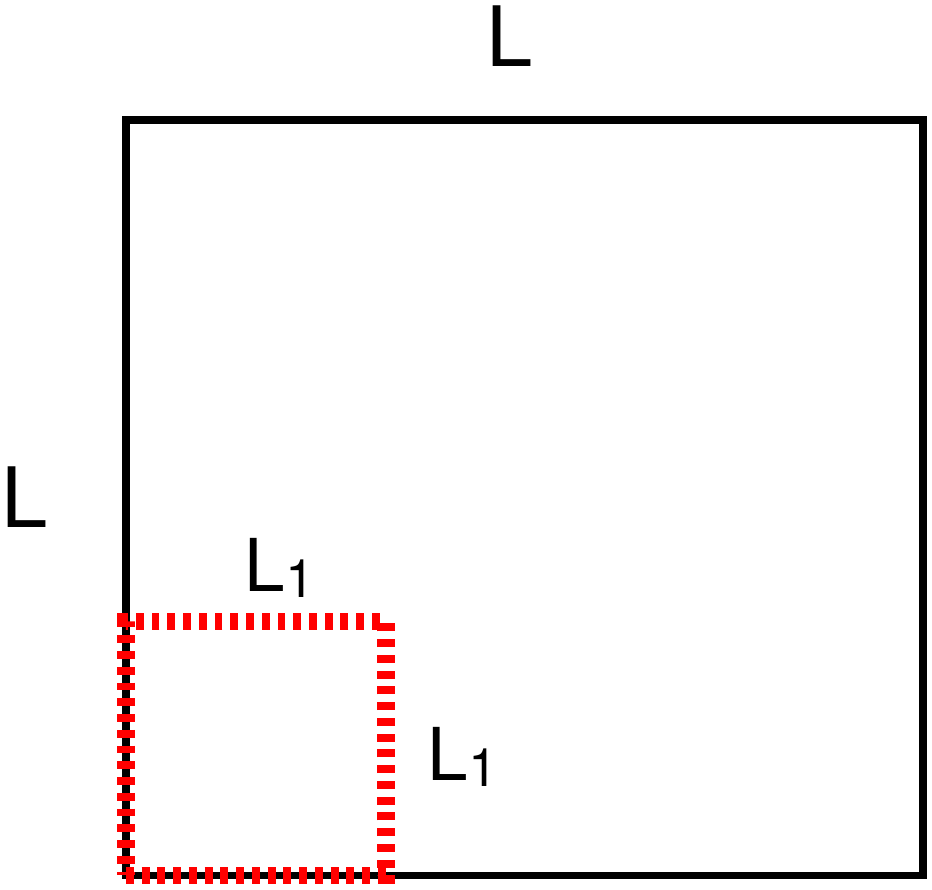}
	\caption{The construction of a configuration for the testing set related to a given linear box size $L_1$ from a spin configuration on a $L$ by $L$ square area. }
	\label{testing}
\end{figure}

For a configuration on a $L$ by $L$ squared area generated from the Monte Carlo simulations, the spins belonging to the first $L_1$ by $L_1$ squared sub-area are considered as one of the testing set for that given $L_1$, see fig.~\ref{testing} and the associated caption for a detailed explanation. For a given $Q$-state configuration with $Q$ being even, the substitution rules 1, 3, 5, ... , $q-1$  $\rightarrow$ -1 and 2, 4, 6, ... , $Q$ $\rightarrow$ 1 are used to create a configuration suitable for the NN prediction. For a given $Q$-state configuration with $Q$ being odd,
the same rule applies for $1, 2, 3, ..., q-1$, and $q \rightarrow 1$ or -1 with equal probabilities.   

For convenience, when the 
NN outcomes are presented in the following, $L$ and $L_1$ will be 
used to stand the linear system sizes of the configurations generated by the Monte Carlo calculations and the configurations of the training set, respectively. Notice the testing set has the same linear system size $L_1$
as that of the training set.

\subsection{The magnitude $R$ of the MLP output vectors}

In the testing stage, if the input configuration is highly similar to the training set, then the output vector is close to $(1,0)$ or $(0,1)$ which would result in $R \sim 1$. On the other hand, if the
input configuration is in strong contrast to the training set, then
the output vector is close to $(0.5,0.5)$ which would lead to $R \sim 1/\sqrt{2}$. Inspired by this, for raw spin configurations from MC simulations with a given $L$, the temperature where the associated $R$ take the value of $\left(1+1/\sqrt{2}\right)/2$ is defined to be the pseudo-critical temperature $T_c(L)$ corresponding to that $L$. This approach in conjunction with certain semi-experimental finite-size formulas lead to accurate determination of the critical temperatures $T_c$ for many different physical models. For a detailed explanation of how this works, see Refs.~\cite{Tan20.1,Tse22,Tse23,Tse241,Tse242,Jia24,Tse25}.  

\section{Numerical Results}

To examine the critical behaviors of the studied models using the NN approach described here, we have conducted large scale Monte Carlo calculations (MC)
using the Swendsen-Wang-Kotecky algorithm \cite{Wan89,Wan90} for $Q=2,3,4,5,$ and 6. For every considered $L$, the simulated spin configurations at many various temperatures $T$ are recorded. 
In the following subsections, the MLP outcomes will be presented.

\subsection{The MLP outcomes related to $Q=2$}

First of all, we would like to examine whether a MLP trained with a fixed $L_1 = 32$ can be applied to compute the pseudo-critical temperatures related to configurations with linear system sizes $L \ge 32$. 

\begin{figure}
	\includegraphics[width=0.5\columnwidth]{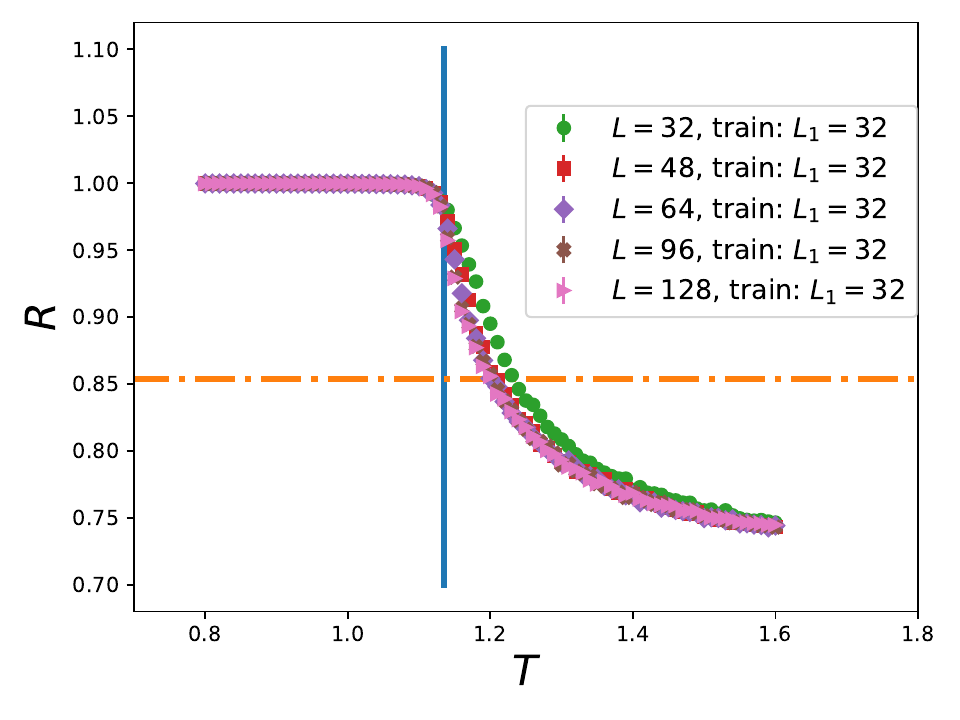}
	\caption{$R$ as functions of $T$ for various $L$. The vertical solid and the horizontal dashed-dotted lines are the expected $T_c = 1.1346$ and $\frac{1+1/\sqrt{2}}{2}$, respectively. The results are determined by a MLP trained with a fixed $L_1 = 32$ using the building units of fig.~\ref{train}. }
	\label{fixedL32}
\end{figure}

$R$ as functions of $T$ for several values of $L$ are shown in fig.~\ref{fixedL32}. The vertical solid and the horizontal dashed-dotted lines in the figure are the expected $T_c = 1.1346$ and
$\frac{1+1/\sqrt{2}}{2}$, respectively. The used $L$ range from 32 to 128. Fig.~\ref{fixedL32} shows that the obtained values of pseudo-critical $T$ stay almost the same for $L \ge 48$. In other words, no useful information regarding the finite-size effects are obtained. This indicates the MLP trained
with a fixed $L_1=32$ is not capable of determining $T_c$ of the $Q=2$
model.   

\begin{figure}
	\includegraphics[width=0.5\columnwidth]{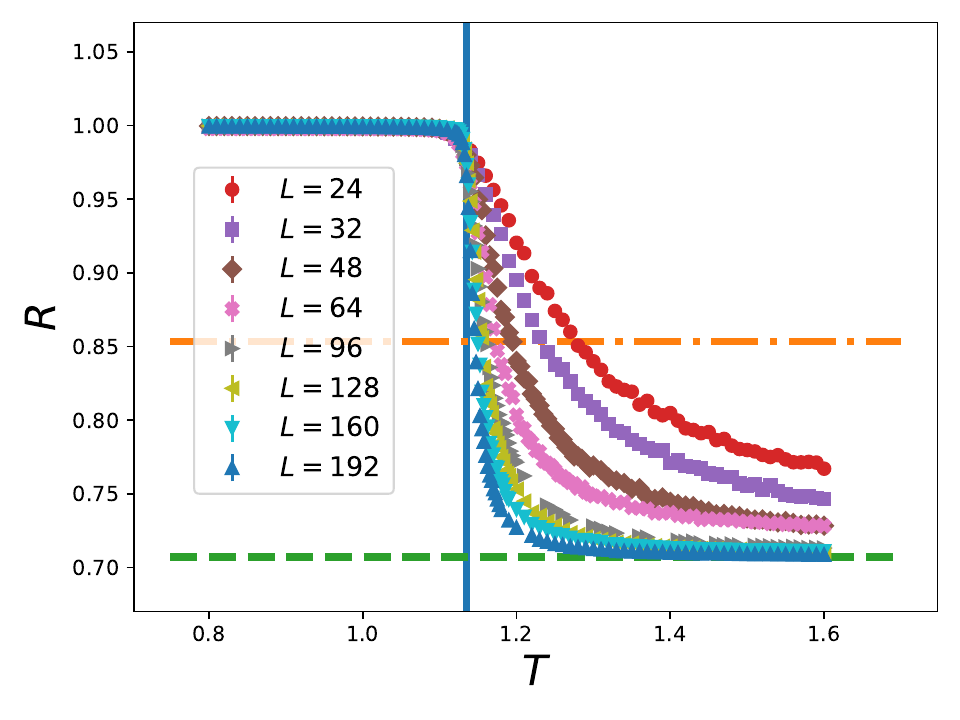}
	\caption{$R$ as functions of $T$ for various $L$. The vertical solid and the horizontal dashed-dotted lines are the expected $T_c = 1.1346$ and $\frac{1+1/\sqrt{2}}{2}$, respectively. For each considered $L$, the results are determined by a MLP trained with a fixed $L_1 = L$ using the building units of fig.~\ref{train}. }
	\label{L1=L}
\end{figure}

Interestingly, if one considers the cases of $L_1 = L$, namely the training set has the same linear system size as that of the raw spin configurations from MC simulations (In this case, the configurations in the testing set are in one to one correspondence with the raw spin configurations), then finite-size effects relevant for carrying out fits to compute the bulk $T_c$ do appear. $R$ as functions of $T$ for several $L$ 
are shown in fig.~\ref{L1=L}. In that figure, the results of $R$
associated with a given $L$ are obtained by a MLP trained with the training set shown in fig.~\ref{train}. In particular, the configurations in the training set(s) have the same linear system size as that of the raw spin configurations from the MC simulations (and the testing set(s)).

As can be seen from fig.~\ref{L1=L}, the pseudo-critical temperatures approach the expected $T_c = 1.1346$ (the vertical solid line in the figure) as $L$ increases. With these pseudo-critical temperatures, one can conduct fits using certain forms to compute the bulk $T_c$, like the ones typically performed in the literature. Here we do not
attempt to carry such calculations since we are more interested in
examining whether our MLP can determine the critical behaviors of
$Q=3,4,5$ and 6 models correctly. The critical behaviors of these models
are intriguing and are more challenge when NN techniques are employed
for the computations.

\begin{figure}
	\includegraphics[width=0.5\columnwidth]{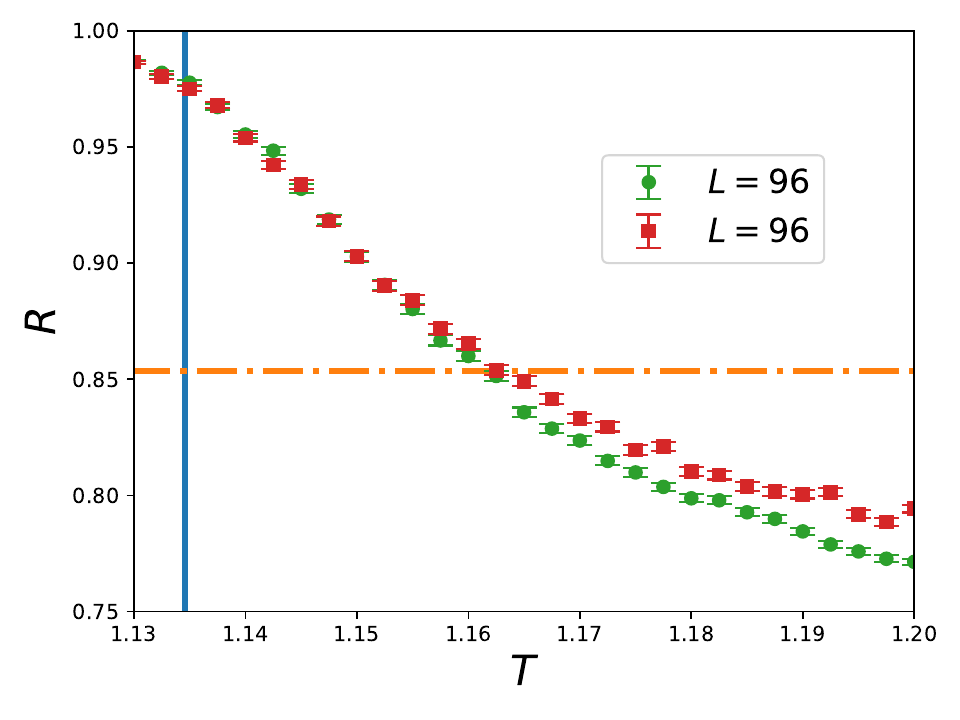}
	\caption{$R$ as functions of $T$ for two different testing sets. The vertical solid and the horizontal dashed-dotted lines are the expected $T_c = 1.1346$ and $\frac{1+1/\sqrt{2}}{2}$, respectively. }
	\label{diff_conf}
\end{figure}

It should be pointed out that even with the same trained MLP employed here, the estimated values of $T_c(L)$ for different testing sets may differ slightly. For example, fig.~\ref{diff_conf}
shows that MLP outcomes of two different testing sets (associated with $Q=2$ model and $L = L_1 = 96$). The horizontal dashed-dotted and vertical solid lines represent $(1+1/\sqrt{2})/2$ and the true $T_c$ of $Q=2$ model, respectively. While the estimated values of $T_c$ for these two different testing sets agree very well, these two data set at some temperatures are off slightly. This indicates the NN-determined $T_c$ may be influenced by the considered testing sets. It has been demonstrated in Ref.~\cite{Tse242} that using different random seeds to train CNNs with the same architecture can lead to statistically different $T_c$. In addition, the number of epochs used also has mild impact on the estimated $T_c$ \cite{Li251}. To conclude, to obtain a NN-calculated $T_c$ (and the related critical exponents) with the most robust error(s) (which take into
account the statistical and systematic uncertainties), several NNs with various random seeds, epochs, batchsize and architectures should be trained. Moreover, different training and testing sets must be considered as well.         

These statistical and systematic effects may also occur for traditional methods such as the Monte Carlo simulations.

\subsection{The NN outcomes related to $Q=3,4,5,6$}

\begin{figure}
	\includegraphics[width=0.5\columnwidth]{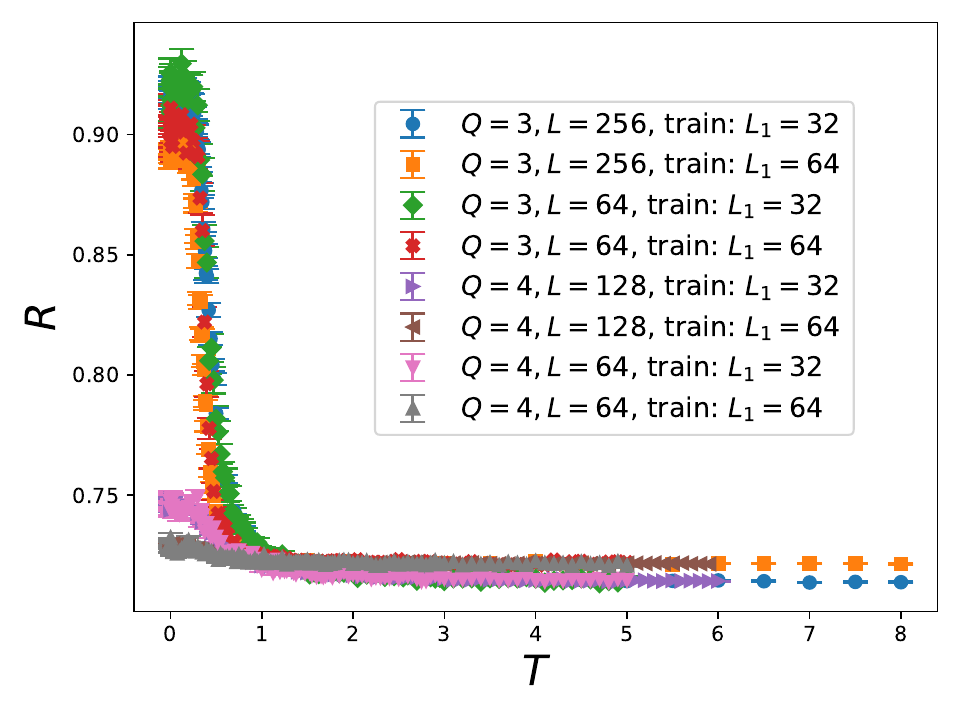}
	\caption{$R$ as functions of $T$ and various $L$ for $Q=3$ and $Q=4$ models. }
	\label{q3q4}
\end{figure}

For each of $Q=3$ and 4 models, two linear system sizes $L$ are considered and
two trainings with $L_1=32$ and $L_1=64$ are performed. 
The obtained MLP outcomes are demonstrated in fig.~\ref{q3q4}.
The results shown in the figure reveal the message that when one moves from high-$T$ region to low-$T$ region, the values of $R$ corresponding to the $Q=3$ model rise from a value near $1/\sqrt{2}$ to a value around 0.9. In addition, for $Q=4$ model, $R$ stay quite close to $1/\sqrt{2}$ even at very low temperatures. 

For the $Q=3$ model, the rising of $R$ from $1/\sqrt{2}$ to 0.9 when one approaches the very low temperature region from the high temperature region indicates that certain kind of order similar to the training set, namely the staggered order, establishes at very low-$T$ region. Moreover, the observation that the values of $R$ quickly decrease to a value near $1/\sqrt{2}$ for $T > 0.25$ (For the case of $L=256$ and $L_1 = 64$) implies
that the system is already in the disordered phase for $T > 0.25$. 
Combining all these outcomes, one concludes that
for $Q=3$ model a phase transition takes place at a very low temperature. This is consistent with the theoretical prediction that
$Q=3$ model has $T_c = 0$. 

For the $Q=4$ model, at high temperature, it is known that the system is disordered and the corresponding configurations are in strong contrast to the training set. Hence, one expects $R \sim 1/\sqrt{2}$
at the high-$T$ region and this is exactly what's found. As one lowers the temperatures, the resulting $R$ still stay close to $1/\sqrt{2}$ even for the case of $T \sim 0.001$. 

For a $Q$-state antiferromagnetic Potts model on the square lattice,
the Potts variable at a site has strong tendency of being different from
those of its nearest neighboring sites if the model is ordered at the low temperature region. Hence, for this case, after the procedure of constructing the testing set from the raw spin configurations, a generated configuration in the testing set should have a pattern similar to the training set. This will lead to a scenario that $R$ will rise from $1/\sqrt{2}$ to a value close to 1 when one moves from the high-$T$ region to the low-$T$ region.

The behaviors of $R$ as functions of $T$ shown in the figure suggest strongly that for $Q=4$ model, no order is established at any finite $T$. This conclusion is in agreement with the theoretical result that
$Q=4$ model is always disordered at any finite $T$.

\begin{figure}
	\includegraphics[width=0.5\columnwidth]{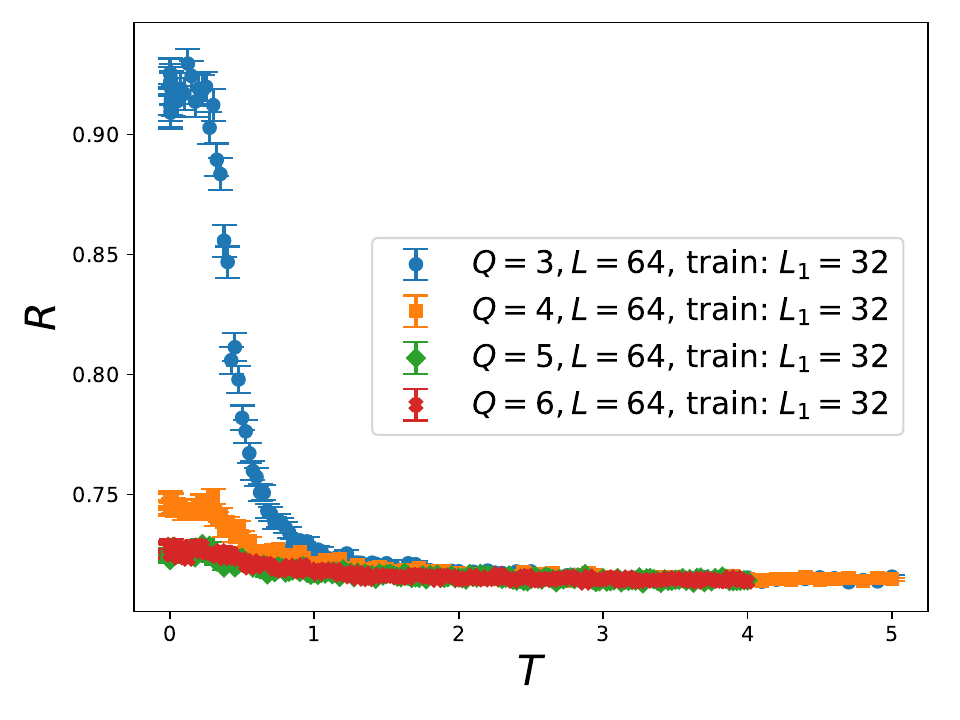}
	\caption{$R$ as functions of $T$ for $L=64$ and $L_1=32$ for $Q=3,4,5$ and 6 models. }
	\label{q3q4q5q6}
\end{figure}

For $Q=5$ and $Q=6$ models, we have carried out the calculations for
$L=64$ and $L_1=32$. The associated results together with those related to $Q=3$ and $Q=4$ models are depicted in fig.~\ref{q3q4q5q6}.
The outcomes in the figure clearly imply strongly that only $Q=3$ model is ordered at very low temperature, and $Q=4,5$ and 6 models 
are disordered at any finite $T$.

Since $Q=3,5$ are odd and $Q=4,6$ are even, namely, at least two even $Q$ and
and two odd $Q$ are considered in this study. Hence, the fact that only the $R$ of 
$Q=3$ model rise to a value close to 1 at the low-$T$ region is a consequence of an     
intrinsic feature of the $Q$-state antiferromagnetic Potts models on the square lattice, not an artificial effect due to our procedure of conducting the NN computations.

\section{Discussions and Conclusions}

In this study, we explore the critical behaviors of 2D antiferromagnetic $Q$-state Potts models for $Q=2,3,4,5$ and 6 on
the square lattice. It is well-known that the $Q=3$ model is ordered only at zero temperature and $Q=4,5,6$ models are always disordered at any (finite) temperature. 

The method of NN approach considered here is adopted from Ref.~\cite{Li25}. In particular, the
used MLP consists of one input layer, one hidden layer of 2 or 512 neurons, and one output layer. In addition, the MLP was trained previously using two artificially made stagger-type configurations in Ref.~\cite{Li25} and no training is conducted here. 

Remarkably, the MLP with such a simple architecture and is trained without any real physical configurations can be applied to compute the critical temperature of the 2D antiferromagnetic 2-state Potts models on the square lattice. It also uncover the critical behaviors
correctly for the 2D antiferromagnetic 3-, 4-, 5-, and 6-state models on the square lattice. Specifically, these models are disordered either at any finite temperature ($Q=3$ model) or even at zero temperature ($Q=4,5,6$ models).

Previously, a MLP with the same architecture was trained by two artificially made ferromagnet-type configurations consisting of 200 sites (using 0 and 1 as the values for these 200 sites). With dedicated way of building the testing set from the raw spin configurations, the obtained MLP can successfully compute the critical points of
several models, such as the three-dimensional (3D) classical $O(3)$
model, the 2D classical $XY$ and the 3-state generalized classical $XY$ models, and the 2D ferromagnetic $Q$-state Potts models with $Q=2,3,4,5,6,8,9,10$. It can determine the critical exponents of the 2- and 4-state ferromagnetic Potts models and the non-equilibrium Ising models with reasonable precision. In summary, this mentioned MLP is
universal to a very large extent. It should be pointed out that, 
using the standard deviations of $R$, the MLP can also identify the
critical temperatures of antiferromagnetic 2-, 3- and 4-state Potts model on the square lattice as well. For the case of $Q=2$ model,
due to the fact that STD is a quantity with small magnitude compared to that of $R$, the computed $T_c$ is less accurate and can be influence by several NN-related parameters. The use of $R$ will suffer milder effects from the mentioned tunable parameters. Hence, the unconventional training approach shown in this study would lead to more stable results than that using the strategy of Ref. \cite{Tan20.1} when
antiferromagnetic systems are considered.

The simple MLP with the training strategy considered here has been applied successfully to study the critical phenomena of 2D antiferromagnetic Ising model with nearest and next-to-nearest interactions. It is demonstrated convincingly that these investigated phase transitions are first order and the associated signals are very persuasive.

\begin{figure}
	\vbox{
		\includegraphics[width=0.5\columnwidth]{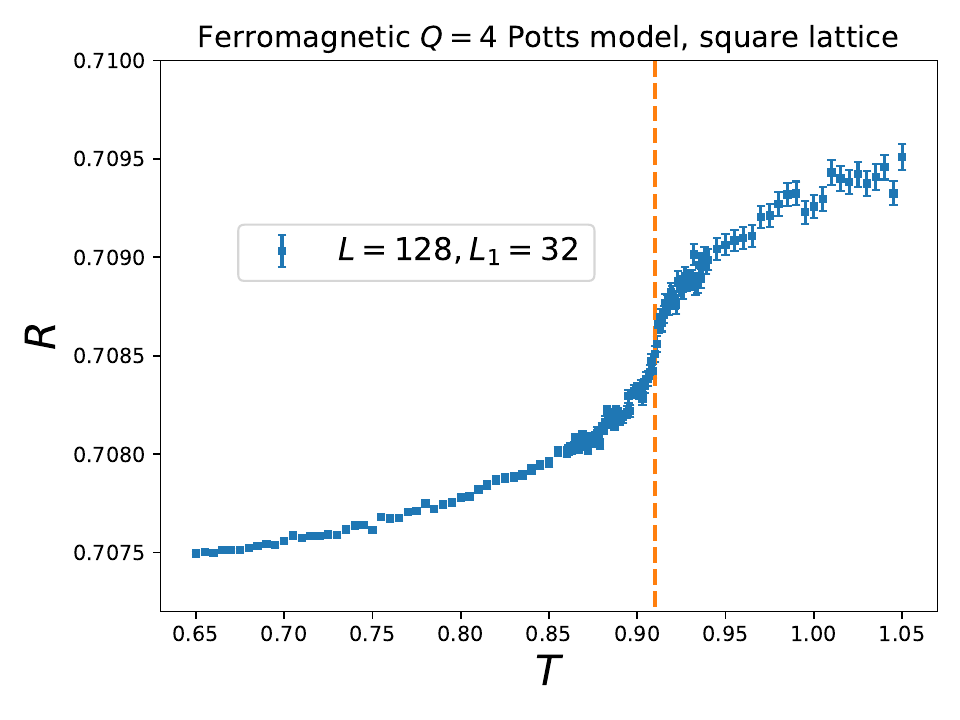}
		\includegraphics[width=0.5\columnwidth]{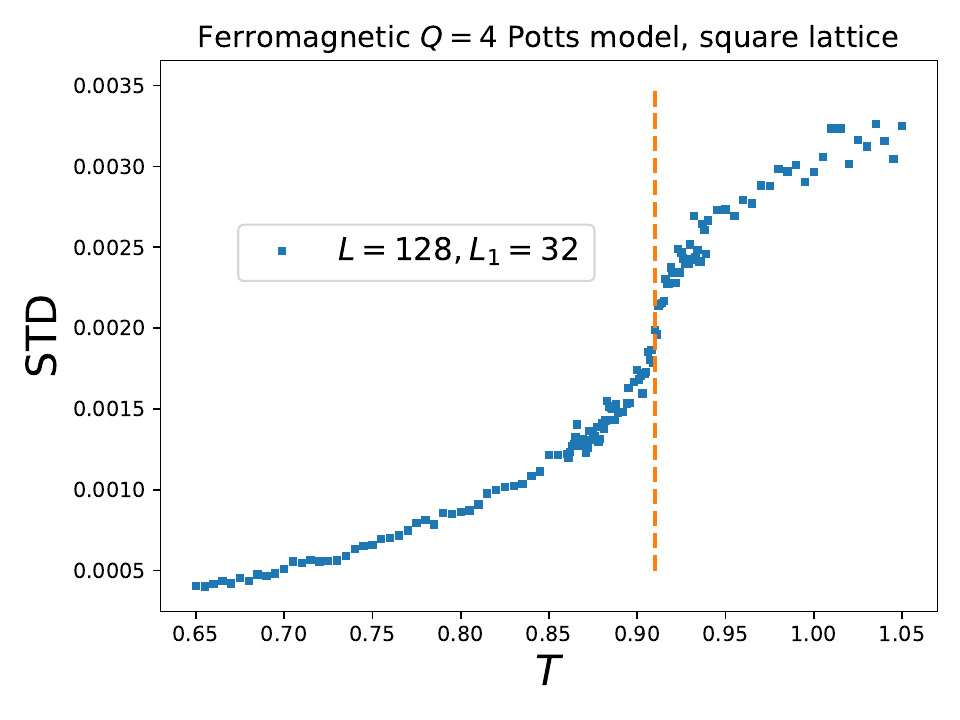}
	}
	\caption{$R$ (top) and their associated standard deviations STD (bottom) as functions of $T$ with $L=128$ and $L_1=32$ for the $Q=4$ ferromagnetic Potts model on the square lattice. }
	\label{f4}
\end{figure}

\begin{figure}
	\vbox{
		\includegraphics[width=0.5\columnwidth]{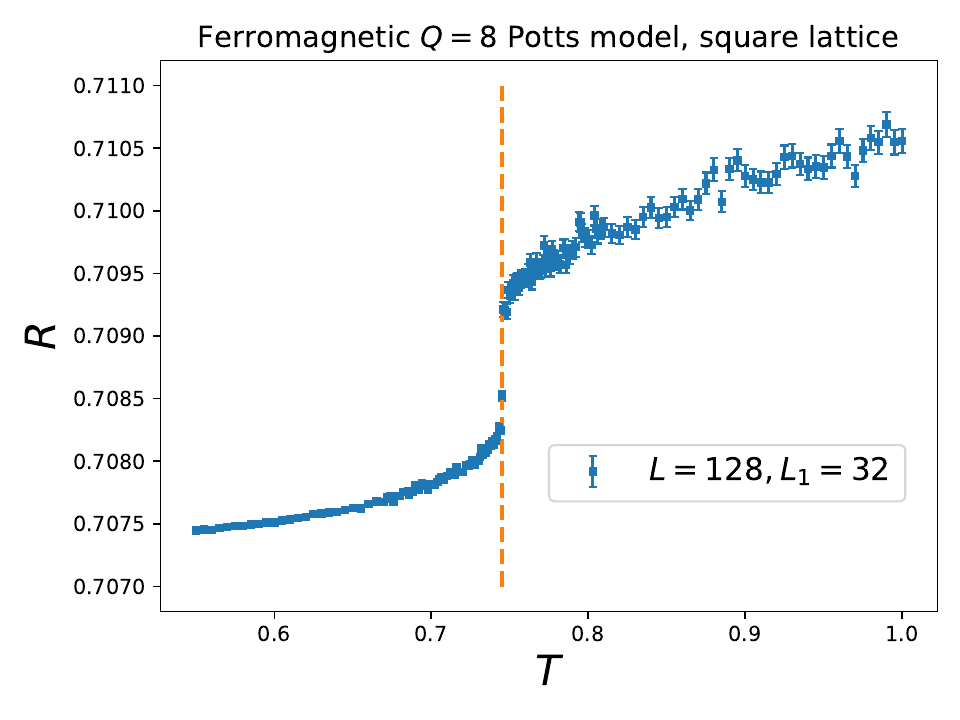}
		\includegraphics[width=0.5\columnwidth]{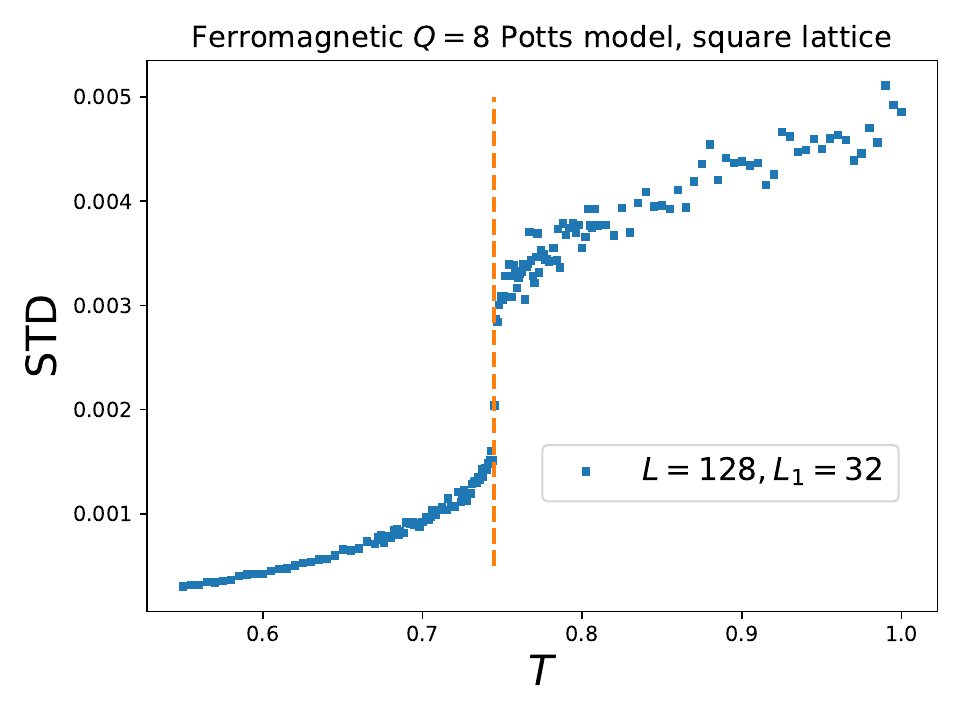}
	}
	\caption{$R$ (top) and their associated standard deviations STD (bottom) as functions of $T$ with $L=128$ and $L_1=32$ for the $Q=8$ ferromagnetic Potts model on the square lattice. }
	\label{f8}
\end{figure}

An immediate examination of the applicability of present MLP trained with the configurations of fig.~\ref{train} is to see whether it can be employed to determine the critical temperatures and the critical behaviors of the $Q$-state ferromagnetic Potts models on the square lattice.  
Figs.~\ref{f4} and \ref{f8} are the associated outcomes with such investigations for the 4-state and 8-state ferromagnetic Potts models on the square lattice.
In all panels of the figs.~\ref{f4} and \ref{f8}, the vertical dashed lines are the expected $T_c$. 

As can be seen from both panels of fig.~\ref{f4}, the curves of $R$ and STD of $R$
turn sharp upwardly around $T \sim 0.9$. This scenario is consistent with the fact that
the value of $T_c$ for the 4-state ferromagnetic Potts model on the square lattice is
$T_c \sim 0.9102$. Moreover, these curves do not show any discontinuity, implying the 
transition is second order. 

The curves of $R$ and STD of $R$ corresponding to the 8-state ferromagnetic Potts model are
shown as the top and the bottom panel of fig.~\ref{f8}, respectively. Apparently, sudden jumps occur exactly at the theoretical $T_c$. In addition, the large gaps shown in both 
panels of the figure suggests convincingly that the phase transition for the 8-state ferromagnetic Potts model is first order. This agrees with the expectation. It should be pointed out that the outcomes given in figs.~\ref{f4} and \ref{f8} are obtained with $L_1 = 32$. When $L_1=64$ is used, only the STD of $R$ reveals the right critical information for these two models. It is likely that STD of $R$ is a more appropriate quantity 
than $R$ when ferromagnetic systems are studied. In any case,
while further studies are needed, the outcomes demonstrated in figs.~\ref{f4} and \ref{f8}
are encouraged.

In this study, the MLP considered correctly identifies the critical behaviors of
the 2D antiferromagnetic 3-, 4-, 5-, and 6-state Potts models on the square lattice.
The outcomes presented here also suggest strongly that only subsets of the whole configurations are sufficient to determine correctly the critical behaviors of the studied models. This is likely applicable to other systems with unusual ground states and phase transitions as well.  
Its applicability to study the phase transitions of the 2D $Q$-state ferromagnetic Potts models on the square lattice is also promising.
In other words, the built and already trained MLP likely has potential to explore other models, particular those with untypical critical phenomena without being retrained. This will be left as a future work.

It should be pointed out that the criterions for determining whether a phase transition is first order for the cases of $L = L_1$ and $L > L_1$ found in (revised version of) Ref.~\cite{Li25} are different. In particular, the NN parameter batch\_size used for $L = L_1$ and $L > L_1$ are 80 and 40, respectively. Interestingly, if one considers batch\_size  $= 40$ for the case of $L = L_1$, then the resulting criterion is the same as the one of $L > L_1$. Both criterions associated with $L = L_1$ identify 
the phase transitions studied in Ref.~\cite{Li25} correctly (First-order phase transitions). It may be interesting to explore this further to see if there is any 
subtlety when the configurations of fig.~\ref{train} are used as the training set(s).    

Finally, we would like to emphasize the fact that because the phase transitions of the models
considered here are well-known based on the order-parameter approach, they are perfectly suited
to examine the universal applicability of our unconventional NN method due to their untypical critical behaviors. The results are promising and further validation by applying the constructed NN to other models is desirable. Such examinations are currently under investigation.

\section*{Aacknowledgments}
The first two authors contributed equally to this work. The MLP models used in this study were implemented using the Keras API of TensorFlow \cite{tens}. Parts of the manuscript also appear in the master’s thesis of the first author (K.W.H., National Taiwan Normal University, Taipei, Taiwan, July 2026).
Partial support from National Science and Technology Council (NSTC) of
Taiwan is acknowledged (Grant numbers: NSTC 113-2112-M-003-014-, NSTC 114-2112-M-003-004-, and 
NSTC 115-2112-M-003-003-).

\section*{Conflict of Interest}

The authors declare no conflict of interest.

\section*{Data Availability Statement}

Data are available from the corresponding author
on reasonable request.

\end{document}